\documentclass[10pt,leqno]{amsart}
\usepackage{graphicx}
\usepackage{indentfirst,csquotes}

\usepackage{amssymb,amsthm,amsmath}
\usepackage{xcolor,paralist,hyperref,titlesec,fancyhdr,etoolbox}

\titlespacing*{\section}{0pt}{3.5ex plus 1ex minus .2ex}{2.3ex plus .2ex}

\makeatletter
\providecommand{\@secnumpunct}{.\enspace}
\makeatother

\hypersetup{ colorlinks=true, linkcolor=black, filecolor=black, urlcolor=black }

\newif\ifisanonymized
\isanonymizedfalse

\usepackage[acronyms,toc,nonumberlist,nogroupskip]{glossaries}
\glsdisablehyper
\glsnoexpandfields

\usepackage{verbatim}
\usepackage{tabularx,calc}
\usepackage{color}
\usepackage{adjustbox}
\usepackage{hyperref}
\usepackage{ifthen}
\usepackage{lipsum}
\usepackage{xstring}
\usepackage{amssymb}
\usepackage{amsmath}
\usepackage{siunitx}
\usepackage{graphicx}
\usepackage{bm}
\usepackage{booktabs}
\usepackage{multirow}
\usepackage{booktabs}
\usepackage{colortbl}
\usepackage{microtype}
\usepackage{here}
\usepackage{tabularx}
\usepackage{arydshln}
\usepackage{titlesec}
\usepackage{array}
\usepackage{siunitx}
\usepackage{listings}
\usepackage{adjustbox}
\usepackage{titlesec}
\usepackage{caption}
\usepackage{xspace}
\usepackage{subcaption}
\usepackage{listings}
\usepackage{multicol}
\usepackage[bibstyle=ieee,
			citestyle=numeric-comp,
			backend=bibtex,
			mincitenames=1,
			maxcitenames=1,
			minbibnames=4,
			maxbibnames=4,
			doi=false,
			isbn=false,
			url=false,
			natbib=true]
			{biblatex}

\DefineBibliographyStrings{english}{andothers={et\addabbrvspace al\adddot}}
\DefineBibliographyStrings{english}{phdthesis = {PhD Thesis,}}
\AtEveryBibitem{\clearlist{language}}
\newcommand{\parencitep}[2]{\parencite[#2]{#1}}
\newcommand{\linquote}[3]{\inq{#1}~\parencitep{#2}{#3}}
\newcommand{\textcitebad}[2]{\textcite{#1}}

\newcommand{\etal}{~et~al.}

\newcommand{\inq}[1]{``#1''}

\renewenvironment{quote}
{\vspace{0.15em}\list{}{\rightmargin=10pt
    \leftmargin=10pt}%
  \item\relax}
{\endlist\vspace{0.15em}}

\newcounter{definitionCounter}

\newcounter{rquestionCounter}

\newcounter{problemCounter}

\newcommand{\BeforeAndAfterRequirementOffset}{0.25em}

\newcounter{requirementCounter}
\newcommand{\requirementPrefix}{REQ}

\newcommand{\setRequirementType}[1]{%
    \renewcommand{\requirementPrefix}{#1}%
    \setcounter{requirementCounter}{0}%
}

\makeatletter
\newcommand{\requirement}[2]{%
    \refstepcounter{requirementCounter}%
    \edef\@tempa{\requirementPrefix}%
    \csdef{reqprefix@#1}{\@tempa}%
    \label{req:#1}%
    \par\vspace{\BeforeAndAfterRequirementOffset}%
    \begin{quote}%
        \noindent\textbf{\@tempa~\therequirementCounter:} #2%
    \end{quote}%
    \vspace{\BeforeAndAfterRequirementOffset}%
}

\newcommand{\reqref}[1]{%
    \expandafter\@reqref\csname reqprefix@#1\endcsname{#1}%
}
\newcommand{\@reqref}[2]{%
    #1~\ref{req:#2}%
}
\makeatother
\newcommand{\ads}{automated driving system\xspace}

\newcommand{\adss}{automated driving systems\xspace}

\newcommand{\ArchExperts}{Architectural System Engineers\xspace}

\newcommand{\asar}{AUTOSAR Adaptive\xspace}

\newacronym{asoa}{ASOA}{Automotive Service-Oriented Architecture}
\newcommand{\asoa}{\gls{asoa}\xspace}

\newcommand{\atool}{\qemph{Architecture Tool}\xspace}

\newcommand{\autocargo}{\qemph{auto}CARGO\xspace}

\newcommand{\autoelf}{\qemph{auto}ELF\xspace}

\newcommand{\autoshuttle}{\qemph{auto}SHUTTLE\xspace}

\newcommand{\autotaxi}{\qemph{auto}TAXI\xspace}

\newcommand{\autotech}{autotech.agil\xspace}

\newcommand{\autox}{\qemph{auto}X\xspace}

\newcommand{\av}{automated vehicle\xspace}
\newcommand{\avs}{automated vehicles\xspace}

\newacronym{fmea}{FMEA}{Failure Mode and Effects Analysis}
\newcommand{\fmea}{\gls{fmea}\xspace}

\newcommand{\FunOwners}{Function Experts\xspace}

\newacronym{gsn}{GSN}{Goal Structuring Notation}

\newacronym{hazop}{HAZOP}{Hazard and Operability Analysis}
\newcommand{\hazop}{\gls{hazop}\xspace}

\newacronym{icd}{ICD}{Interface Control Document}

\newacronym{icds}{ICDs}{Interface Control Documents}

\newacronym{json}{JSON}{JavaScript Object Notation}

\newacronym{mrm}{MRM}{minimum risk maneuver}

\newcommand{\pii}{\pind interface\xspace}

\newcommand{\piis}{\pind interfaces\xspace}

\newcommand{\PIIExperts}{\SelfP Coordinators\xspace}

\newcommand{\piimechs}{\pii mechanisms\xspace}

\newcommand{\piireqs}{\pii requirements\xspace}

\newacronym{pilog}{PI Log}{performance indicator log}

\newcommand{\pilog}{\gls{pilog}\xspace}

\newcommand{\PiLogLong}{Performance Indicator Log\xspace}

\newacronym{pi}{PI}{performance indicator}
\newcommand{\pind}{\gls{pi}\xspace}

\newcommand{\pinds}{\glspl{pi}\xspace}

\newcommand{\qemph}[1]{\emph{#1}}

\newcommand{\qvk}{quality vector}

\newacronym{ros}{ROS~2}{Robot Operating System 2}
\newcommand{\ros}{\gls{ros}\xspace}

\newcommand{\sad}{self-adaptive\xspace}

\newcommand{\sady}{self-adaptivity\xspace}

\newcommand{\SafeEngineers}{Safety Engineers\xspace}

\newcommand{\saw}{self-aware\xspace}

\newcommand{\sawa}{self-awareness\xspace}

\newcommand{\SelfP}{Self-Perception\xspace}
\newcommand{\Selfp}{Self-perception\xspace}
\newcommand{\selfp}{self-perception\xspace}

\newacronym{servsa}{ServSA}{Service Self-Assessment}

\newacronym{soa}{SOA}{service-oriented architecture}
\newcommand{\soa}{\gls{soa}\xspace}
\newcommand{\soas}{\glspl{soa}\xspace}

\newacronym{spi}{SPI}{Safety Performance Indicator}
\newcommand{\spi}{\gls{spi}\xspace}
\newacronym{spis}{SPIs}{Safety Performance Indicators}
\newcommand{\spis}{\gls{spis}\xspace}

\newacronym{stpa}{STPA}{System-Theoretic Process Analysis}
\newcommand{\stpa}{\gls{stpa}\xspace}

\newacronym{syssa}{SysSA}{System Self-Assessment}

\newcommand{\unicar}{UNICAR\emph{agil}\xspace}

\usepackage[none]{hyphenat}
\titleformat{\paragraph}[runin]{\normalfont\normalsize\itshape}{\theparagraph}{0.5em}{\hspace{0.5em}}[:]
\titlespacing*{\paragraph}{0pt}{1ex plus 0.5ex minus 0.2ex}{0.5em}
\renewcommand{\theparagraph}{\alph{paragraph})}

\begin{document}

% 1. Title
\begin{center}
{\Large\bfseries Coordinating Stakeholders in the Consideration of Performance Indicators and Respective Interface Requirements for \\ Automated Vehicles}
\end{center}

\vspace{1em}

% 2. Technical Report
\begin{center}
\textit{Technical Report}
\end{center}

\vspace{1em}

% 3. Authors and Affiliations
\begin{center}
Richard Schubert$^{1}$,
Marvin Loba$^{1}$,
Alexander Bl\"odel$^{2}$,\\
David Kl\"uner$^{3}$,
Alexandru Kampmann$^{3}$,
and Steven Peters$^{2}$\\[0.5em]
{\small
$^{1}$Institute for Control Engineering, TU Braunschweig, Braunschweig, Germany\\
$^{2}$Institute of Automotive Engineering, TU Darmstadt, Darmstadt, Germany\\
$^{3}$Chair of Embedded Software, RWTH Aachen University, Aachen, Germany\\[0.3em]
\texttt{richard.schubert@tu-bs.de}, \texttt{m.loba@tu-bs.de}, \texttt{alexander.bloedel@tu-darmstadt.de},\\
\texttt{kluener@embedded.rwth-aachen.de}, \texttt{kampmann@embedded.rwth-aachen.de}, \texttt{steven.peters@fzd.tu-darmstadt.de}
}
\end{center}

\vspace{1em}

% 4. Abstract
\begin{center}
\textbf{ABSTRACT}
\end{center}
\noindent
This paper presents a process for coordinating stakeholders in their consideration of performance indicators and respective interface requirements for automated vehicles. These performance indicators are obtained and processed based on the system's \selfp and enable the realization of \saw and \sad vehicles. This is necessary to allow SAE Level 4 vehicles to handle external disturbances as well as internal degradations and failures at runtime. Without such a systematic process for stakeholder coordination, architectural decisions on realizing \selfp become untraceable and effective communication between stakeholders may be compromised. Our process-oriented approach includes necessary ingredients, steps, and artifacts that explicitly address stakeholder communication, traceability, and knowledge transfer through clear documentation. Our approach is based on the experience gained from applying the process in the \autotech project, from which we further present lessons learned, identified gaps, and steps for future work.

\let\thefootnote\relax
\footnotetext{This research was accomplished as part of the projects
\autotech (FKZ~16EMO0285) and
\unicar (FKZ~01IS22088R).
We acknowledge the financial support for both projects by
the Federal Ministry of Research, Technology, and Space (BMFTR; formerly Federal Ministry of Education and Research of Germany, BMBF).}

\bigskip

\section{Introduction}
At SAE~Level~4~\parencite{sae_j3016_2021}, \avs must be able to operate safely without driver supervision. Uncertainty during vehicle operation presents a major challenge here: External disturbances, internal degradations, and failures (e.g., of sensors, actuators, or mechanical components) can never be fully avoided~\parencite{nolte_supporting_2020, nolte_werte_2024}. To address this challenge, the paradigms of \qemph{\sawa} and \qemph{\sady} are particularly promising: \sad systems, relying on their \sawa, shall assess their own capabilities, including their performance, and derive behavioral decisions accordingly, especially under such aforementioned conditions~\parencite{gregory_selfaware_2016, nolte_supporting_2020, schubert_literature_2025}.

For that, \qemph{\selfp} is a prerequisite: \Selfp denotes the process of generating knowledge about the system's internal state through model-based data and information processing, relying on both proprioceptive (internal) and exteroceptive (external) measurements~\parencite[p.~6]{nolte_supporting_2020}. Note that these are not the only capabilities/properties of a system that raise the need for \selfp; others might include, for instance, long-term fleet-level performance monitoring and product enhancement~\parencite{stefansicklinger_how_2022}. The need for \selfp arising from realizing \saw/\sad systems is, however, the motivation of this work.

\Selfp with respect to the system's performance relies on the acquisition of defined \qemph{\pinds} at runtime. The selection of \pinds is crucial for architecture preparation since, e.g., software and hardware interfaces, sensors and processing units must be selected and installed accordingly, even early in system design~\parencite{nolte_towards_2017}.
With the objective of realizing systems with \selfp in mind, we acknowledge current trends in \adss' architectures: On the one hand, increasingly AI-based\footnote{AI $=$ artificial intelligence} architectures and implementations are finding their way into \ads design. Especially (pure) end-to-end approaches raise questions about the need for more atomic functions and implementations~\parencite{tampuu_survey_2022, chen_endtoend_2024, chib_recent_2024}.
On the other hand, modularized solutions are still necessary due to a higher level of transparency, support of decentralized development, and partial update- and/or replaceability~\parencite{vector_serviceoriented_2017, mckinsey_whenmasteringcode_2021}. As an (automotive) example, \qemph{automotive middlewares} abstract from software and hardware components by establishing an \qemph{\soa} with modular \qemph{services}. Examples include \asar~\parencite{autosar_adaptive_2024}, the \ros~\parencite{openrobotics_ros_2024}, and the \asoa~\parencite{kampmann_dynamic_2019, kampmann2023dynamic}. As presented in~\parencite{schubert_performance_2025}, a framework for \sad \adss relying on \pinds---assembled in a so-called \qemph{\qvk}---to be exchanged between services in the \asoa is presented. In this work, we further focus on \soas, and acknowledge their ability to incorporate also AI-based implementations of individual services.

Once \pinds have been selected, interfaces that allow obtaining \pinds need to be defined and harmonized---given possibly already existing (preliminary) architectures. This integration effort directly poses the questions of careful communication and documentation of architectural decisions required due to \pii integration in a general process framework.
This proved to be also a major challenge in the \autotech~\parencite{vankempen_autotechagil_2023, vanKempen2025autotechAgil} project that focused, among others, on the advancement of the four \autox\footnote{\autox is a placeholder for the four vehicle prototypes \autoelf, \autoshuttle, \autotaxi, and \autocargo that share the same architecture.} vehicle prototypes from the \unicar project~\parencite{woopen_unicaragil_2018} for diverse use cases in an urban environment. In the project, among diverse groups of (project) stakeholders, communication difficulties remained a central challenge. For the very specific problem of \pii, based on experience, it was often unclear which specific information was required in the context of \pii by respective stakeholders.

In line with literature, we find that the absence of a systematic process in multi-stakeholder environments leads to fragmented knowledge, untraceable architectural decisions, and ineffective communication---especially in disruptive technologies~\parencite{loba_showcasing_2024, NolteRTAF_Beck25}.
As a solution, in this paper, we present a process for the systematic coordination of stakeholders in their consideration of \piireqs arising from the specific need for realizing \saw and \sad \avs and that was co-developed during the \autotech project, explicitly designed to establish transparent communication channels, ensure full traceability of decisions, and facilitate knowledge transfer across diverse stakeholder groups.

In the remainder of this paper, we present related work and working terminology in \autoref{sec:related} and  \autoref{sec:terminology}. Building on this, in order to thoroughly describe our process, we present our initial process assumptions and derive requirements in \autoref{sec:requirements}. We present the process including relevant roles and activities in \autoref{sec:process}. Finally, having applied the process in the research project \autotech, we reflect on lessons learned and outline gaps as well as the need for future work in \autoref{sec:conclusion}.
\section{Related Work} \label{sec:related_work} \label{sec:related}
With respect to \selfp, in SAE~J3016~\parencite{sae_j3016_2021}, the need for monitoring of \qemph{vehicle} and \qemph{driving automation system performance} is expressed, referring to activities that aim to continuously evaluate the \av's behavior during operation and prepare countermeasures when the dynamic driving task is not realized as intended~\parencite[p.~16-17]{sae_j3016_2021}. For \avs without human supervision, \textcite{maurer_flexible_2000} stresses that functions must provide quality measures for performance monitoring by the system itself at runtime. Building on~\parencite{reschka_fertigkeiten_2017}, \textcite{nolte_towards_2017} propose capability graphs to decompose, refine, and assign safety requirements (with associated measures) to capabilities and later functions of an \av, supporting their monitoring.

\Selfp concepts have increasingly been integrated into industrial \adss: Zoox, for example, introduced a \qemph{robot monitor} intended to maintain an overall assessment of vehicle health across hardware and software layers. This monitoring framework shall extend conventional diagnostic functions by enabling fault detection and coordination among redundant subsystems~\parencite{zoox_introducing_2021}. Comparable principles are applied by Alphabet’s Waymo, as discussed in~\parencite{waymo_waymo_2021}.
Especially in such an industry context, we assume that the selection of \pinds and respective architectural decisions follows defined processes. However, such processes are, to the best of our knowledge, not yet publicly available.

Among published sources, on the other hand, we find a plethora of approaches towards the systematic derivation of a set of \pinds to be measured and---but yet again without the rigorous notion of a process:

\textcite{hawkins_identifying_2023} elaborate on an approach for systematically deriving monitoring requirements from an analysis of a safety case. A safety case shall provide a structured argument for demonstrating that a system is sufficiently safe under specified assumptions~\parencite{koopman_ul4600_2023}. To \inq{measure} the validity of such a safety case, it is recommended to assign \spis~\parencite{koopman_ul4600_2023} to goals/claims, in order to assess whether underlying assumptions hold~\parencite{koopman_ul4600_2023, hawkins_identifying_2023}. Hence, the need to obtain \pinds might further arise in order to provide/disprove evidence for \qemph{safety cases} at runtime.

Further contributions include the approach of \textcite{asaadi_dynamic_2020}, who present the concept of \emph{Dynamic Assurance Cases} that rely on continuous runtime evaluation of assurance properties. \textcite{reich_engineering_2020} also discuss approaches for runtime safety monitors of cyber-physical systems using \qemph{Digital Dependability Identities}---a tool to link elements of a safety argument with architectural elements.

\textcite{gautham_stpadriven_2022} provide another framework for systematically deriving monitors for technical systems using \stpa~\parencite{leveson_engineering_2011} as a framework. By investigating given hazards, identifying \qemph{causal factors} for so-called \qemph{loss scenarios}~\parencite{leveson_engineering_2011}, and relying on a formalization of, e.g., the context and assumptions, dedicated \qemph{data}, \qemph{network}, and \qemph{functional monitors} are derived. %  to monitor \qemph{unsafe data}, \qemph{control paths}, and \qemph{control actions}, respectively.
% The approach hence is similar to those mentioned earlier: Artifacts of a safety-driven design (i.e., a hazard log) lead to a systematic definition of \pii requirements.

The work of \textcite{perez_monitoring_2022} shows an approach towards the automatic derivation of monitoring nodes within an \ros-based framework directly from defined requirements, in which specific quantities are constrained by threshold conditions and multiple such conditions are concatenated. The proposed toolchain automatically associates dedicated \ros \qemph{monitoring nodes} with such requirements, which collect and supervise the corresponding quantities at runtime.

Overall, we find diverse needs for specifying \pinds to measure in the aforementioned literature that can be traced back, at least implicitly, to selected stakeholder perspectives, and desired artifacts; even without an explicit elaboration. Nonetheless, the plethora of approaches raises the question of how they can be consolidated and harmonized in a systematic process. Such a process is necessary to deal with conflicts in (or prior to) architectural decisions. However, the authors are not aware of any such explicit process perspective that is publicly available.
Building on earlier work from our group~\parencite{loba_showcasing_2024} that we use for methodological guidance, defining a process for the selection of \pinds to measure and respective architectural decisions to make is hence the subject of this paper hereafter.
%
% To provide a path towards such, this paper's approach and overall structure is greatly inspired by the work of \textcite{loba_showcasing_2024}. In their paper, a systematic release process for automated vehicle prototypes is presented, aiming to improve communication and management of risks when exposing prototypes to public traffic. The authors propose an incremental approval process for public demonstrations that defines requirements, process decisions, and responsibilities of different stakeholders.
\section{Key Working Terminology} \label{sec:terminology}
Runtime performance assessment with respect to \selfp is the key subject of this work.
% The capabilities of a system shall come with a performance measure expressing the system's current conditions~\parencite{wasson_system_2015, nolte_werte_2024}.
We refer to \textcitebad{wasson_system_2015}{Wasson\etal} who define performance as \linquote{a quantitative measure characterizing a physical or functional attribute relating to the execution of an operation or function}{wasson_system_2015}{5}, i.e., it is a measure of \inq{how well} a system operates~\parencite{wasson_system_2015}. % Furthermore, the \qemph{performance level} is defined as \linquote{an objective, measurable parameter that serves to bound the ability of a system to perform a function [...].}{wasson_system_2015}{28}.
We further use the term \qemph{performance indicator} (PI) to refer to a measurable quantity that characterizes the performance of a system (element).
A \qemph{\pii} is the architectural interface through which a \pind shall be provided.
% 
% Performance levels can be separated into a \qemph{nominal} and \qemph{available} (or \qemph{current}~\parencite{schubert_performance_2025}) performance level~\parencite{stolte_taxonomy_2021}. The two might differ due to the occurrence of degradations and/or 
%
The performance might degrade in case of degradations and/or failures~\parencite{stolte_taxonomy_2021}. In accordance with ISO~26262~\parencite{international_organization_for_standardization_iso_2018}, both degradation and failure originate from a fault but with different outcome, i.e., performance is reduced (degradation) or functionality is lost (failure; see~\parencite{stolte_taxonomy_2021}).
\section{Process Assumptions \& Requirements}  \label{sec:requirements}
% \section{On the Overarching Development Process} \label{sec:process}
%
Inspired by~\parencite{loba_showcasing_2024}, in the following, we outline our process assumptions, requirements, and provide an overview of the derived process.
%
% \subsection{Assumptions and Roles}
%
\begin{figure}%[!h]
    \centering
    \includegraphics[width=0.8\columnwidth]{figures/process.pdf}
    \caption{Scheme of an iterative, systematic design process for \adss, taken from~\parencite{graubohm_systematic_2017}, based on~\parencite{Maurer2002}.}
    \label{fig:procsource}
\end{figure}
%
% In the context of this work, we discuss the domain of \ad and focus on \sawadss, such as the \autox vehicles. 
%
% QUELLEN EINFACH VON MARVIN ÜBERNOMMEN
For the overarching system development process, we assume an iterative approach as shown in \autoref{fig:procsource}~\parencite{graubohm_systematic_2017}: Indicated through its closed, iterative topology, prior prototype or product generations act as intermediate outcomes, potentially providing reusable knowledge and legacy components. Thus, we consider an existing preliminary \ads architecture hereafter. As we focus on \piireqs, in particular, the process described hereafter is conducted \qemph{alongside} this overarching system development process that addresses multiple needs (far beyond establishing \piis for \selfp, i.e., it is not our objective to replace such an overarching process and/or share the same comprehensiveness). Instead, it participates in the overall process during the requirements elicitation, domain-specific design, and integration stages in \autoref{fig:procsource}.
\begin{figure*}%[!t]
    \centering
    \includegraphics[width=0.95\textwidth,trim=5.9cm 1.1cm 5.9cm 1.1cm,clip]{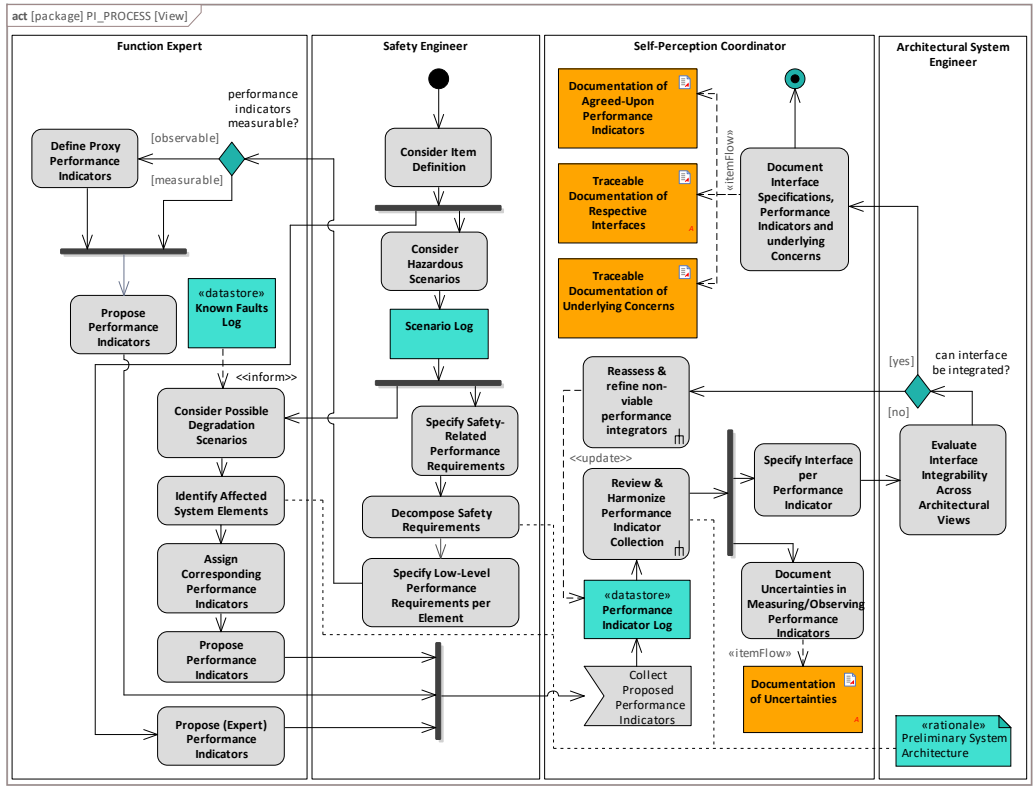}
    \caption{SysML-1.5-based \qemph{Activity Diagram} of the proposed process and relevant roles, created in \qemph{EnterpriseArchitect}. \qemph{Start/end points} are marked by big dots. Boxes indicate \qemph{SysML activities}, bars indicate \qemph{forks/joins}, and \inq{diamonds} indicate decisions. Blue elements indicate early process objects (``ingredients''); orange elements indicate process artifacts, i.e., outcomes.\vspace{-0.4cm}}
    \label{fig:activity-diagram}
\end{figure*}

% \section{Process Requirements} \label{sec:requirements}
%
\setRequirementType{PRO-REQ}
%
% ANMERKUNG LOBA: Ehrlicherweise wären das schon fast Anforderungen/Eigenschaften, die ich durch den Prozessentwurf fur SP- Realization erreichen will? -- RS: BEWUSST ERSTMAL DAGEGEN ENTSCHIEDEN, WEIL ICH EIN BISSCHEN WENIGER RICHTUNG FORM SONDERN EHER RICHTUNG INPUT-OUTPUT ARGUMENTIEREN MÖCHTE
In order to handle \piireqs systematically, collaboratively, and transparently, requirements are derived for the process. First, before going over to the actual process activities, a crucial requirement recognized in~\parencite{loba_showcasing_2024} is to define clear roles to which activities are later assigned:
\requirement{roles-clear}{The process shall comprise clearly defined roles to which activities are assigned.}
%
% As considered in this paper, \sawadss shall monitor their own performance over short time horizons and respond appropriately. Hence, when taking the specific perspective on \piireqs hereafter, we consider those that, from our experience, are particularly relevant for this specific scope. However, we intend that other perspectives, e.g., long-term fleet-level performance monitoring for product enhancement, could be easily integrated into the process.
% 
% Accordingly, before defining more formal requirements, we outline our intent: The final process should (1) start from a defined item (the \autox), (2) consider the need for acquiring \pinds from various perspectives, (3) consolidate the collection of \pinds, (4) and establish respective interfaces for accessing these \pinds within the architecture.
%
For the process activities, their \inq{start/end points} shall be defined. As a minimum prerequisite and hence starting point, the process requires an item definition---comprising, as assumed here, the preliminary system architecture, use cases, and hence possible scenarios:
\requirement{start-itemdefinition}{The process shall start from the item definition.}
Next, we define the desired outcome of our process: In architectural design, well-defined interfaces are fundamental for consistent communication within complex systems and for enabling effective monitoring~\parencite{kurtev_runtime_2017, davies_interface_2020}. Interfaces must be formalized and standardized, e.g., by means of an \qemph{Interface Control Document} (ICD), defining (among others) exchanged data, semantics, and timing~\parencite{davies_interface_2020}. This is also acknowledged in, e.g., the \asoa framework, where service interfaces are described independently of implementation details~\parencite{kampmann_dynamic_2019, kampmann_asoa_2022, schubert_performance_2025} as well as in the extended interface description format \qemph{S$^2$I$^2$} by \textcite{klamann_ansaetze_2024} for modular systems.
We define:
\requirement{end-interfaces}{The process shall yield a set of defined \piis as its major artefact.}
%
% Note that different kinds of systems exist: \qemph{White-box} systems allow traceability, while \qemph{black-box} systems limit direct interpretation of internal behavior~\parencite{garces_whitebox_2016, kotonya_componentbased_2008}. Furthermore, legacy systems may lack the structured interfaces required for monitoring, whereas newly developed architectures can integrate them from the outset~\parencite{allman_managing_2012}. This is why, as acknowledged later in \autoref{sec:process}, meaningful interface definitions should come with an assessment of their integrability.
%
Next, according to the related work in \autoref{sec:related}, safety-related concerns and corresponding artifacts of a safety lifecycle (such as a safety argumentation) are considered a major driver of \piireqs: % (from the scope of this work):
\requirement{topdown-safety}{The process shall consider \piireqs identified systematically from a top-down, safety-related perspective.}
Broadly speaking, performance monitoring usually focus on the performance of the system as a whole and how it can be observed systematically using available \pinds (e.g., as in the capability monitoring approaches in~\parencite{reschka_fertigkeiten_2017, schubert_prototypical_2024}), with the goal of matching them against safety-related performance requirements.
Besides this top-down perspective, bottom-up approaches exist. For instance, fault-centric monitoring approaches (e.g., as in~\parencite{klar_symptom_2011}) may focus on (low-level) faults and how they lead to degradations and/or failures of the system, ultimately impacting the system's performance as a whole.
%
% In general, a bottom-up perspective might open up a view for
Furthermore, besides more systematic methodologies, the fact that intuitive insights from domain experts (here: \FunOwners) can be valuable~\parencite{obrien_developing_1985} has long been recognized. Recent insights further imply that systematic approaches often suffer from implicit assumptions and/or missing information that experts may identify~\parencite{harel_methodical_2023}.
This might specifically be the case for legacy (sub-)systems, where domain expertise is required---but insufficient documentation and/or knowledge transfer channels may exist \parencite{comella-dorda_survey_2000}.
% Apart from the technical domain, for instance, a case study conducted in hospitals implied that experts can contribute to the development of meaningful strategies through an iterative, consultative process with managers as they can act as \inq{stewards} who flexibly adapt their expertise to the needs of other actors in the system~\parencite{liff_experts_2021} (here: \PIIExperts, \SafeEngineers, and \ArchExperts).
%
Fusing both perspectives is generally advised, e.g., also in~\parencite{ISO26262}. We therefore define:
\requirement{bottomup-fault}{The process shall consider \piireqs identified systematically from a bottom-up, safety-related perspective.}
%
% Besides more systematic methodologies, the fact that intuitive insights from domain experts (here: \FunOwners) can be valuable~\parencite{obrien_developing_1985} has long been recognized. Recent insights further imply that systematic approaches often suffer from implicit assumptions and/or missing information that experts may identify~\parencite{harel_methodical_2023}.
% Apart from the technical domain, for instance, a case study conducted in hospitals implied that experts can contribute to the development of meaningful strategies through an iterative, consultative process with managers as they can act as \inq{stewards} who flexibly adapt their expertise to the needs of other actors in the system~\parencite{liff_experts_2021} (here: \PIIExperts, \SafeEngineers, and \ArchExperts).
% To open our process for such experts, we define:
%
% \requirement{bottomup-expert}{The process shall consider \piireqs identified from a bottom-up, expert-driven perspective.}
%
Once \piireqs have been identified (at least in a preliminary form), their accessible documentation is required in order to be able to discuss and refine them collaboratively. Here, \qemph{accessible} means that the documentation is \inq{technically} easily accessible to all stakeholders, it is transparent, and can be (to the best of the process engineers' knowledge) easily understood by all stakeholders. Furthermore, it should be traceable, i.e., reviewers can trace back the origin of the documentation to the respective considerations (e.g., a safety-related requirement):
\requirement{arch-decisions-doc}{The process shall explicitly cover the accessible and traceable documentation of artifacts.}
Lastly, we assume an iterative approach to be suitable for the multi-step reconsideration of \piireqs between different perspectives and stakeholders. We define:
\requirement{iterative-process}{The process shall contain iterative steps, i.e., key artifacts are revisited and refined in feedback loops involving all relevant stakeholders.}
Note that this also implies that the formats chosen to fulfill PRO-REQ~6 support reiteration of architectural documentations in an accessible way as well.

\section{Process Overview} \label{sec:process}
The following process roles and activities contribute to the realization of architectures that fulfill specific \piireqs. They are further visualized in an \qemph{SysML Activity Diagram} in \autoref{fig:activity-diagram}. Activities are associated with specific roles (in swimlanes). They are described as follows.
\subsection{Process Roles}
As required by PRO-REQ~1, we define the following roles: % process roles that, in our opinion, span the \inq{dimensions} of our intended process meaningfully:

\paragraph*{\PIIExperts}
\PIIExperts represent the central entity for coordinating stakeholders in their consideration of \pinds and respective interface requirements. It may be a single person or group of experts tasked with orchestrating, discussing, and documenting all matters concerning \pii in this work.

\paragraph*{\SafeEngineers}
For the scope of our work, \SafeEngineers are considered to be responsible for selecting relevant scenarios for safety-related analyses as well as defining, documenting, and communicating safety-related (performance) requirements.

\paragraph*{\ArchExperts}
\ArchExperts possess a detailed understanding of the overall system architecture. In plain terms, they shall be able to explain \inq{what is connected to what and how.} Their tasks include defining, documenting, communicating knowledge of (preliminary) system elements, as well as coordinating integration activities.
% As stated in our assumptions, they can rely on a preliminary architecture with legacy components.

\paragraph*{\FunOwners}
\FunOwners are responsible for individual system functions. Their domain-specific expertise ensures that monitoring mechanisms reflect the actual characteristics and constraints of each function. Note that under the term \FunOwners, we also subsume those that ultimately implement the respective functions in software and hardware and possess the corresponding knowledge. In plain terms, \FunOwners should know: \inq{What is the function doing and how, what are its limitations, and how can it fail?}. % They should also know unintended operation of a function can be identified and what (already established) ways to identify such cases are.

\subsection{Process Activities} \label{sec:activities}
%\subsection{Process ``Ingredients'' and Outcome}
%
The starting point is the item definition for the system, allowing to elaborate on relevant use cases and thus scenarios for the system. Also, a preliminary architecture is present (PRO-REQ~2 fulfilled). The shared outcome is the definition of interfaces that enable the provision and exchange of \pinds (PRO-REQ~3 fulfilled).
%
% \subsection{Process Activities}
%
In line with our requirements, between the start and end points, both a top-down and bottom-up perspective are taken by us to define meaningful activities.

\subsubsection{Top-Down Perspective}
The first approach concerns safety-related aspects from a top-down perspective (fulfilling PRO-REQ~4), beginning with the formulation of a set of safety requirements with associated \pinds. Such requirements might be embedded into one or more safety concepts (e.g., for functional~\parencite{ISO26262} or behavioral safety~\parencite{waymo_road_2017}) and/or a safety argumentation, where (in the latter case) the associated \pinds are also referred to as \qemph{\spis} (see \autoref{sec:related}. In particular, safety concepts rely on the decomposition of requirements and thus \pinds. Similarly, safety argumentations rely on the decomposition of safety claims and their substantiation with corresponding evidence. This hierarchical refinement (more or less explicitly) influences the granularity of \spis. In our view, determining the (appropriate) level of granularity thus constitutes a design decision: Same as for the classical safety requirements defined in safety concepts, the obtained value for such \spi might be directly measurable or must be observed for a collection of low-level indicators. Here, hierarchical structures in the argumentation (such as those defined through Goal Structuring Notation) support this decomposition.

In either case, the definition of \pinds here directly leads to the identification of \piireqs. However, the question of the direct measurability at runtime arises, which is an inherently implementation-dependent question and thus requires the involvement of \FunOwners, who may propose the use of \qemph{proxy-measures} to observe the desired \pinds at runtime.

\paragraph*{Example} Consider the example of the SAE Level 4 vehicle prototype \autox \ads that shall operate in urban traffic~\parencite{woopen_unicaragil_2018, vankempen_autotechagil_2023}. For the scenario of a crosswalk with a pedestrian crossing (example taken from~\parencite{bagschik_systems_2018}), relevant capabilities cover the reliable detection of pedestrians. An associated hazard may reflect the possible collision with a pedestrian in case of a missed detection. \SafeEngineers might state that the probability of misdetection is hence a critical \pind to be monitored at runtime.
As this quantity cannot be measured directly, \FunOwners may propose the use of \qemph{proxy-measures} derived from (pseudo-)uncertainty estimation techniques in the AI-based implementation of the perception model, whose outputs shall act as \pinds. These include, e.g., ensemble-based and dropout methods~\parencite{lakshminarayanan_simple_2017, labach_survey_2019}. Also, as shown by \textcite{kuznietsov_methodology_2024}, perception quality further depends on measurable external factors such as time of day, weather, and vehicle speed. Such factors also act as \pinds as they allow observing current perception performance.

\subsubsection{Bottom-Up Perspective}
The second, complementary perspective follows a bottom-up logic (fulfilling PRO-REQ~5). Here, \FunOwners provide the domain-specific knowledge necessary to identify relevant \pinds, considering also known failures and degradations of the respective function. In our experience, \FunOwners understand well how degradations or failures affect the behavior of individual functions and how unintended operation can be detected. \FunOwners may use systematic approaches for fault identification (and impact thereof) such as an \fmea; see~\parencite{ISO26262, iec_60812_2018} for further details. Other methods trace the impact of a fault on the system behavior, e.g., as applied in an \stpa \parencite{leveson_engineering_2011}; see~\parencite{leveson_engineering_2011, thomas_stpa_2015, abdulkhaleq_using_2017, graubohm_identifikation_2023} for further details on using \stpa. Similarly, the established \hazop provides a guideword-based approach to identify deviations from intended functionality~\parencite{kletz2018hazop} in this context. For a comparison of methods covering also \stpa and \fmea, please consider~\parencite{sun_comparison_2021}.
% 
% In this context, even simple (binary) fault indicators can be considered as \pinds, as they allow observing the performance level of the system (element)---and if a fault is relevant to this performance level, then that measure shall allow direct observability. From this point of view, all perspectives can be connected under the concept of \pinds.

\paragraph*{Example} Parallel to the top-down analysis, \FunOwners examine the system from a bottom-up perspective, starting from potential failures and degradations. First of all, cases of total processing unit failure that would not allow the provision of performance information might require \inq{heartbeat} monitoring on redundant hardware. In terms of degradations, examples include sudden overexposure due to direct sunlight, partial occlusion of the lens by dirt or, say, a plastic bag, or performance degradation caused by overheating. The experts therefore propose additional \pinds that can reveal such degradation. One is a binary flag indicating whether the perception function is currently (fully) impaired. Another \pind might concern the temperature.
% 
% \vspace{2mm}
% 
% As noted in \autoref{sec:requirements}, expert inputs provide an important, third perspective for identifying \piireqs (fulfilling PRO-REQ~6). Due to their inherently subjective nature, no further investigation is conducted here.

\subsection{Harmonization of the \PiLogLong}
Different perspectives may yield different \pinds to be monitored. We propose their collection in a \qemph{\pilog}. Within this log, a review and harmonization of the proposed \pinds is required to ensure consistency and avoid redundancies. This action is conducted by the \PIIExperts, in a collaborative manner with other stakeholders. To promote stakeholder discussion, each \pind in the log is expected to come with (at least) a textual description, technically possible value ranges, and a unit. If such a description is lacking or inconsistent, the \PIIExperts shall request additional information from the respective stakeholders. Also, different stakeholders may propose the same (or almost the same \pind), which shall be identified and dropped/merged. The desired outcome is a consolidated \pilog.
For all consolidated \pinds, it is from our experience beneficial to also add information about the uncertainty of the \pind. This information is typically provided by the \FunOwners and documented accordingly.

Note that while the process focuses on identifying and harmonizing \pinds as well as establishing corresponding interfaces, the definition of thresholds for these measures---i.e., determining when a component or module exhibits misbehavior based on the observed \pind values---is considered out of scope for this work. Note further that any later definition of such thresholds would still require \piis in the first place. % , e.g., derived using our process.
% Such threshold definitions are typically part of the monitoring mechanisms that utilize the established \piis and depend on domain-specific safety requirements, acceptable performance levels, and operational constraints.

\subsection{Definition \& Documentation of \pind Interfaces}
For each \pind, a respective interface for acquiring/exchanging the \pind shall be considered and integrated into the preliminary architecture. This action is conducted by the \ArchExperts, in a collaborative manner with other stakeholders, especially the \PIIExperts. Most importantly, the question of whether the interface can \qemph{technically} be implemented must be considered.
% (given a sound definition of the respective \pind).
The questions span multiple architectural layers: at the functional level (provided and consumed information), at the software level (data formats), and at the hardware level (acquisition, transmission, processing with, e.g., bandwidth, timing, and spatial constraints).

While the full answer to this question is inherently implementation-dependent and beyond the scope of this brief process overview, we note that in-depth discussions between \ArchExperts and \FunOwners must be conducted to find a feasible solution---especially if initial integration attempts fail. If interfaces cannot be integrated, this directly leads to a loss of information with respect to the corresponding \pinds that cannot be observed at runtime. Respective reassessment and refinement activities for non-viable \pinds lead hence to their own, larger activity of conflict resolution in the process that, for the sake of readability, is not further disentangled in \autoref{fig:activity-diagram}. % but would, again, involve the aforementioned roles.

For all other interfaces where integration is successful, the \ArchExperts shall integrate the interfaces into the preliminary architecture. As each interface maps to a \pind in the \pilog as well as to the explicit connection of activities in \autoref{fig:activity-diagram}, it becomes clear that architectural interfaces can be traced back to \qemph{agree-upon} \pinds (in the finalized \pilog), their origin (i.e., the perspective taken and the responsible stakeholders), and uncertainty information.

For documentation, different formats might be chosen to document the process artifacts (e.g., simple text files, tables, or more detailed systems-engineering formats). In the \autotech project, besides simple tables during early draft stages, a specific \atool was used to document architectural interfaces and their semantics even in early stages (see \autoref{sec:conclusion}).

Lastly, the process acknowledges the need for iterative feedback by respective stakeholders along its way (fulfilling PRO-REQ~7): Issues during the harmonization of the \pilog and the definition and documentation of architectural interfaces shall be resolved iteratively by the \PIIExperts in collaboration with the respective stakeholders as indicated in \autoref{fig:activity-diagram} and documented alongside.
\section{Lessons Learned, Conclusion, \& Future Work} \label{sec:conclusion}
The described process has been applied and iteratively clarified for the design of \sad \autox~\parencite{woopen_unicaragil_2018} vehicles in the \autotech~\parencite{vankempen_autotechagil_2023} project. The \autotech project brought together around 40 different institutions from industry, science, and the public sector~\parencite{vankempen_autotechagil_2023}. For automation of vehicle systems, the \asoa framework with its modular services was used.

\subsection{Lessons Learned}
As stated in the motivation of this work, communication difficulties proved to be a central challenge. Starting from a clearly defined item and using tools such as (informal) scenario descriptions, a common base for initial discussion could be successfully established.

Furthermore, prior to relying on the developed process, it was often unclear which specific information was required in the context of \pii by the stakeholders. While \SafeEngineers are usually very aware of the need for and the role of \piis (for safety-driven matters), \FunOwners often felt the additional burden of providing respective information. Here, in our experience, explicit consideration of stakeholder needs helped to understand their origin and both clarify and motivate the overall process approach.

Prior to applying the process in the project, the separation of responsibilities with respect to specifying \piireqs was often unclear. Here, a separation between the aforementioned roles defined in our process descriptions has proven to be beneficial. Moreover, the separation of clear roles was further amplified by the definition of distinct \qemph{work packages} in the project, e.g., for various functional development activities, architectural design, system safety, and \pii activities---with respective experts working therein. When expert input was needed, due to the clear work package structure, they could easily be identified and consulted.

To promote the desired process outcome of specified \pinds and respective interfaces, \PIIExperts organized dedicated workshops to discuss and specify them together with \SafeEngineers and \ArchExperts. In these workshops, all roles were present and clearly defined. Confirming our intuition in the process description in \autoref{sec:process}, \FunOwners could more easily follow the bottom-up perspective in \autoref{sec:process} while \SafeEngineers could more easily follow the top-down approach.

%%% ATOOL
%
A major challenge that we underestimated in the beginning of the project was the understanding of the system architecture for the \autox vehicles among all stakeholders and, accordingly, the documentation later in the process (see PRO-REQ~8): As existing prototypes from the previous \unicar project~\parencite{woopen_unicaragil_2018} with their given (legacy) architectures were used, new project members from diverse domains had to familiarize themselves with the systems. Here, we particularly found an accessible documentation of the system architecture, i.e., its system elements and their interfaces, to be of great help. Moreover, we learned that architectural decisions (such as defining \piis) must not only be formally specified but also presented in a way that promotes effective exchange between diverse stakeholders---especially in agile development as in the \autotech projects~\parencite{vankempen_autotechagil_2023}.

Hence, alongside this process, the use of the so-called \atool~\parencite{kampmann_asoa_2022, kampmann2023dynamic} has further been applied in the project to specify the evolving \soa of the \autox vehicles: Originally introduced in the \unicar project and further extended in the \autotech project, the \atool represents a web-based system-engineering platform (i.e., it is accessible via a simple web browser) developed to support the \asoa framework implemented with respect to the aforementioned challenges~\parencite {kampmann2023dynamic}.
During the project(s), using the tool, architectural interfaces with a connection to \piireqs could be successfully specified and documented for various functions. Additional text fields allowed to make comments, e.g., to facilitate understanding of the defined \piis or even provide some (qualitative notion of) traceability. Also, uncertainties could be annotated textually.

However, more rigorous/formal mechanisms for ensuring traceability with respect to the specific focus of our work are lacking: In the design of the \atool, a trade-off between accessibility and feature-richness had to be made. While, in our opinion, the \atool avoids the steep learning curve of conventional modelling tools, it lacks the formal traceability of architectural decisions that model-based engineering environments ensure, such as explicit linkage between requirements, interfaces, and other artifacts. Accordingly, a fully traceable documentation of architectural decisions was not yet achieved.

\subsection{Conclusion \& Future Work}
Based on the lessons learned and the experience gained during the project(s), we conclude that the process described in this work can be a valuable tool for the systematic coordination of stakeholders in their consideration of \piireqs. Overall, we found the process to contribute the necessary clarification during development activities and facilitate the collaborative design of \piimechs.
Note that the considered perspectives justifying \piireqs represent only one specific selection. In the future, we aim to bring in additional perspectives, e.g., fleet-level performance monitoring as mentioned earlier.

% AB: "Die Idee mit dem Prozess finde ich gut. Nur kommt die Erfahrung aus Autotech später als Verifikation rein, obwohl sie ja doch eher der anlass ist. Das müsste ja eher das sein, woraus anforderungen abgeleitet werden. Zusätzlich bspw. dazu, dass Normen das Vorgehen zur Entwicklung nicht weiter spezifizieren und dann in der Praxis nicht klar ist, woher dann die informationen kommen sollen." -- NOTIZ FÜR DIE RELATED WORK MITGENOMMEN
In previous work conducted as part of the \autotech project, reaching further than the subject of \selfp or \sawa, we discussed \sad capabilities of \avs~\parencite{schubert_performance_2025}. Accordingly, \sady of the \autox, i.e., internal and/or external behavior adaptions as a response to insufficient performance levels, were explicitly discussed and specified in additional workshops. While we found the respective discussions to be valuable, we were not able to find a trivial way to integrate the systematic derivation of \qemph{adaptation actions} \parencite{schubert_literature_2025} into the process described in this work. Hence, future work should further investigate the integration of \sady into the process if proven to be beneficial.
\section{Acknowledgement}
\ifisanonymized
% Anonymized version: show black box with same height
\rule{\columnwidth}{\AnonymizedAcknowledgementHeight}
\else
% Normal version: show actual acknowledgement
We would like to thank our partners from the \autotech project for the stimulating discussions, participation in workshops, and overall application of the process.
\fi

\printbibliography

@misc{sae_j3016_2021,
  title = {J3016 -- {{Taxonomy}} and {{Definitions}} for {{Terms Related}} to {{On-Road Motor Vehicle Automated Driving Systems}}},
  author = {{SAE}},
  year = {2021, Accessed: Oct. 29 2025},
  urldate = {2025-07-16},
}

@article{nolte_supporting_2020,
  title = {Supporting {{Safe Decision Making Through Holistic System-Level Representations}} \& {{Monitoring}} -- {{A Summary}} and {{Taxonomy}} of {{Self-Representation Concepts}} for {{Automated Vehicles}}},
  author = {Nolte, Marcus and Jatzkowski, Inga and Ernst, Susanne and Maurer, Markus},
  year = {2020},
  journal = {arXiv:2007.13807},
}

@phdthesis{nolte_werte_2024,
  title = {Werte- Und F{\"a}higkeitsbasierte {{Bewegungsplanung}} F{\"u}r Autonome {{Stra{\ss}enfahrzeuge}} -- {{Ein}} Systemischer {{Ansatz}}},
  author = {Nolte, Marcus},
  year = {2025},
  school = {Techn. Univ. Braunschweig},
  doi = {10.24355/dbbs.084-202501271119-0},
  note = {\href{https://doi.org/10.24355/dbbs.084-202501271119-0}{doi:~10.24355/dbbs.084-202501271119-0}},
  address = {Braunschweig, Germany}
}

@misc{gregory_selfaware_2016,
 author = {Gregory, Irene M and Leonard, Charles and Scotti, Stephen J},
 note = {Report},
 title = {Self-aware vehicles: {Mission} and performance adaptation to system health},
  year = {2016},
 urldate = {2025-08-21},
}

@misc{schubert_literature_2025,
  title = {{Architectural Requirements for Self-Aware and Self-Adaptive Automated Driving Systems: A Literature Review}},
  author = {Schubert, Richard},
  year = {2026},
  note = {unpublished}
}

@misc{stefansicklinger_how_2022,
  title = {How the {{Big Loop}} Powers Data-Driven Development for {{ADAS}}/{{AD}}},
  author = {{Stefan Sicklinger}},
  year = {2022, Accessed: Oct. 29 2025},
  journal = {CARIAD},
  urldate = {2025-03-24},
  url = {https://cariad.technology/de/en/news/stories/big-loop-introduction.html},
  note = {\href{https://cariad.technology/de/en/news/stories/big-loop-introduction.html}{https://cariad.technology/de/en/news/stories/big-loop-introduction.html}},
}

@inproceedings{nolte_towards_2017,
  title={Towards a skill-and ability-based development process for self-aware automated road vehicles},
  author={Nolte, Marcus and Bagschik, Gerrit and Jatzkowski, Inga and Stolte, Torben and Reschka, Andreas and Maurer, Markus},
  booktitle = {Proc. IEEE 20th Int. Conf. Intell. Transp. Syst. (ITSC)},
  pages={1--6},
  year = {2017},
  organization={IEEE},
  doi = {10.1109/ITSC.2017.8317814},
  note = {\href{https://doi.org/10.1109/ITSC.2017.8317814}{doi:~10.1109/ITSC.2017.8317814}},
}

@article{tampuu_survey_2022,
  title = {A {{Survey}} of {{End-to-End Driving}}: {{Architectures}} and {{Training Methods}}},
  shorttitle = {A {{Survey}} of {{End-to-End Driving}}},
  author = {Tampuu, Ardi and Matiisen, Tambet and Semikin, Maksym and Fishman, Dmytro and Muhammad, Naveed},
  year = {2022},
  journal = {IEEE Trans. Neural Networks Learn. Syst.},
  volume = {33},
  number = {4},
  pages = {1364--1384},
  doi = {10.1109/TNNLS.2020.3043505},
  urldate = {2025-08-22},
  note = {\href{https://doi.org/10.1109/TNNLS.2020.3043505}{doi:~10.1109/TNNLS.2020.3043505}},
}

@article{chen_endtoend_2024,
  title = {End-to-{{End Autonomous Driving}}: {{Challenges}} and {{Frontiers}}},
  shorttitle = {End-to-{{End Autonomous Driving}}},
  author = {Chen, Li and Wu, Penghao and Chitta, Kashyap and Jaeger, Bernhard and Geiger, Andreas and Li, Hongyang},
  year = {2024},
  journal = {IEEE Trans. Pattern Anal. Mach. Intell.},
  volume = {46},
  number = {12},
  pages = {10164--10183},
  doi = {10.1109/TPAMI.2024.3435937},
  urldate = {2025-08-22},
  note = {\href{https://doi.org/10.1109/TPAMI.2024.3435937}{doi:~10.1109/TPAMI.2024.3435937}},
}

@article{chib_recent_2024,
  title = {Recent {{Advancements}} in {{End-to-End Autonomous Driving Using Deep Learning}}: {{A Survey}}},
  shorttitle = {Recent {{Advancements}} in {{End-to-End Autonomous Driving Using Deep Learning}}},
  author = {Chib, Pranav Singh and Singh, Pravendra},
  year = {2024},
  journal = {IEEE Trans. Intell. Vehicles},
  volume = {9},
  number = {1},
  pages = {103--118},
  doi = {10.1109/TIV.2023.3318070},
  urldate = {2025-08-22},
  note = {\href{https://doi.org/10.1109/TIV.2023.3318070}{doi:~10.1109/TIV.2023.3318070}},
}

@misc{vector_serviceoriented_2017,
  title = {Service-Oriented {{Architectures}} and {{Ethernet}} in {{Vehicles}}},
  author = {{VECTOR}},
  year = {2017, Accessed: Oct. 31 2025},
  url = {https://cdn.vector.com/cms/content/know-how/_technical-articles/PREEvision/PREEvision_SOA_Ethernet_ElektronikAutomotive_201703_PressArticle_EN.pdf},
  urldate = {2025-07-16},
  note = {\href{https://cdn.vector.com/cms/content/know-how/_technical-articles/PREEvision/PREEvision_SOA_Ethernet_ElektronikAutomotive_201703_PressArticle_EN.pdf}{https://cdn.vector.com/cms/content/know-how/\_technical- articles/PREEvision/PREEvision\_SOA\_Ethernet\_ElektronikAutomot ive\_201703\_PressArticle\_EN.pdf}},
}

@misc{mckinsey_whenmasteringcode_2021,
  title = {Mastering Automotive Software},
  author = {{McKinsey}},
  year = {2021, Accessed: Nov. 02 2025},
  urldate = {2025-01-08},
  url = {https://www.mckinsey.com/industries/automotive-and-assembly/our-insights/when-code-is-king-mastering-automotive-software-excellence},
  note = {\href{https://www.mckinsey.com/industries/automotive-and-assembly/our-insights/when-code-is-king-mastering-automotive-software-excellence}{https://www.mckinsey.com/industries/ automotive-and-assembly/ our-insights/when-code-is-king-mastering-automotive-software- excellence}},
}

@misc{autosar_adaptive_2024,
  title = {Adaptive {{Platform}} of {{AUTOSAR}}},
  author = {{AUTOSAR}},
  year = {2024, Accessed: Nov. 01 2025},
  journal = {AUTOSAR},
  urldate = {2025-01-07},
  url = {https://www.autosar.org/standards/adaptive-platform},
  note = {\href{https://www.autosar.org/standards/adaptive-platform}{https://www.autosar.org/standards/ adaptive-platform}},
}

@misc{openrobotics_ros_2024,
  title = {{{ROS}} -- {{Robot Operating System}}},
  author = {{OpenRobotics}},
  year = {2024, Accessed: Oct. 22 2025},
  journal = {ROS},
  urldate = {2025-01-07},
  url = {https://ros.org/},
  note = {\href{https://ros.org/}{https://ros.org/}},
}

@inproceedings{kampmann_dynamic_2019,
	title = {A {Dynamic} {Service}-{Oriented} {Software} {Architecture} for {Highly} {Automated} {Vehicles}},
	copyright = {All rights reserved},
	doi = {10.1109/ITSC.2019.8916841},
  booktitle = {Proc. IEEE Intell. Transp. Syst. Conf.},
	author = {Kampmann, Alexandru and Alrifaee, Bassam and Kohout, Markus and Wüstenberg, Andreas and others},
	year = {2019},
	pages = {2101--2108},
  note = {\href{https://doi.org/10.1109/ITSC.2019.8916841}{doi:~10.1109/ITSC.2019.8916841}},
}

@PHDTHESIS{kampmann2023dynamic,
  author = {Kampmann, Alexandru},
  othercontributors = {Kowalewski, Stefan and Eckstein, Lutz},
  title = {{A} dynamic service-oriented software architecture for the automotive domain},
  school = {RWTH Aachen University},
  year = {2023},
  doi = {10.18154/RWTH-2024-00864},
  note = {\href{https://doi.org/10.18154/RWTH-2024-00864}{doi:~10.18154/RWTH-2024-00864}},
}

@misc{schubert_performance_2025,
  title = {{{Performance}} {{Assessment}} and {{Management}} in {{Service-Oriented Architectures}} for {{Automated Driving Systems}}},
  author = {Schubert, Richard and others},
  year = {2026},
  note = {unpublished}
}

@inproceedings{vankempen_autotechagil_2023,
  title = {{{AUTOtech}}.Agil: {{Architecture}} and {{Technologies}} for {{Orchestrating Automotive Agility}}},
  booktitle = {Proc. 32th Aachen Colloq. Sustain. Mobility},
  author = {van Kempen, Raphael and Lampe, Bastian and Leuffen, Marc and Wirtz, Lena and others},
  date = {2023-10-10},
  year = {2023},
  note = {\href{https://doi.org/10.18154/RWTH-2023-09783}{doi:~10.18154/RWTH-2023-09783}}
}

@inproceedings{woopen_unicaragil_2018,
  title = {{{UNICARagil}} -- {{Disruptive}} {{Modular Architectures for Agile, Automated Vehicle Concepts}}},
  booktitle = {Proc. 27th Aachen Colloq.},
  author = {Woopen, Timo and Lampe, Bastian and Böddeker, Torben and Eckstein, Lutz and others},
  year = {2018},
  doi = {10.18154/RWTH-2018-229909},
  note = {\href{https://doi.org/10.18154/RWTH-2018-229909}{doi:~10.18154/RWTH-2018-229909}},
}

@book{NolteRTAF_Beck25,
  author ={Nolte, Marcus and Braun, Niklas and Fleischer, Torsten and Kolarova, Viktoriya and Loba, Marvin and Salem, Nayel Fabian and Steege, Hans and Form, Thomas and Stolte, Torben and H\"ohmann and Poszler, Franziska and Reiff, Tobias and Hense, Benno and Maurer, Markus},
  title = {Anmerkungen zu {Sicherheit} und {Risiken} autonomer {Straßenfahrzeuge} -- {Teil} 1 \& 2},
  publisher = {{C.H.} {BECK}},
  note = {NZV -- Neue Zeitschrift für Verkehrsrecht, NZV 5/2025 (p.~198-207), continued in NZV 6/2025 (p.~241-251)}
}

@phdthesis{maurer_flexible_2000,
    author = {Markus Maurer},
    title = {{{Flexible}} {{Automatisierung}} von {{Straßenfahrzeugen}} mit {{Rechnersehen}}},
    school = {Universität der Bundeswehr},
    address = {Munich, Germany},
    year = {2000}
}

@phdthesis{reschka_fertigkeiten_2017,
  title = {Fertigkeiten- Und {{F{\"a}higkeitengraphen}} Als {{Grundlage}} Des Sicheren {{Betriebs}} von Automatisierten {{Fahrzeugen}} Im {\"O}ffentlichen {{Stra{\ss}enverkehr}}},
  author = {Reschka, Andreas},
  year = {2017},
  school = {Techn. Univ. Braunschweig},
  address = {Braunschweig, Germany}
}

@misc{zoox_introducing_2021,
  title = {Introducing {{Zoox Safety Innovations}} -- {{Safety Report}}},
  author = {{Zoox}},
  year = {2021, Accessed: Oct. 31 2025},
  url = {https://www.datocms-assets.com/106048/1696536139-zoox_safety_report_volume2_2021_v2.pdf},
  urldate = {2025-07-14},
  note = {\href{https://www.datocms-assets.com/106048/1696536139-zoox_safety_report_volume2_2021_v2.pdf}{https://www.datocms-assets.com/106048/1696536139-zoox\_safety report\_volume2\_2021\_v2.pdf}},
}

@misc{waymo_waymo_2021,
  title = {{{Waymo}} {{Safety}} {{Report}}},
  author = {Waymo},
  year = {2021, Accessed: Oct. 26 2025},
  url = {https://downloads.ctfassets.net/sv23gofxcuiz/4gZ7ZUxd4SRj1D1W6z3rpR/2ea16814cdb42f9e8eb34cae4f30b35d/2021-03-waymo-safety-report.pdf},
  urldate = {2025-07-14},
  note = {\href{https://downloads.ctfassets.net/sv23gofxcuiz/4gZ7ZUxd4SRj1D1W6z3rpR/2ea16814cdb42f9e8eb34cae4f30b35d/2021-03-waymo-safety-report.pdf}{https://downloads. ctfassets.net/sv23gofxcuiz/4gZ7ZUxd 4SRj1D1 W6z3rpR/2ea16814cdb42f9e8eb34cae4f30b35d/2021-03-waymo- safety-report.pdf}},
}

@inproceedings{hawkins_identifying_2023,
	address = {Cham},
	title = {Identifying {Run}-{Time} {Monitoring} {Requirements} for {Autonomous} {Systems} {Through} the {Analysis} of {Safety} {Arguments}},
	doi = {10.1007/978-3-031-40923-3_2},
	booktitle = {Proc. Comput. Saf., Rel., Secur.},
	publisher = {Springer Nature Switzerland},
	author = {Hawkins, Richard and Ryan Conmy, Philippa},
	editor = {Guiochet, Jérémie and Tonetta, Stefano and Bitsch, Friedemann},
	year = {2023},
	pages = {11--24},
  note = {\href{https://doi.org/10.1007/978-3-031-40923-3_2}{doi:~10.1007/978-3-031-40923-3\_2}},
}

@article{koopman_ul4600_2023,
  title = {{{UL}} 4600: {{What}} to Include in an Autonomous Vehicle Safety Case},
  author = {Koopman, Philip},
  year = {2023},
  journal = {Comput.},
  volume = {56},
  number = {05},
  pages = {101--104}
}

@article{asaadi_dynamic_2020,
	title = {Dynamic {Assurance} {Cases}: {A} {Pathway} to {Trusted} {Autonomy}},
	volume = {53},
	copyright = {https://ieeexplore.ieee.org/Xplorehelp/downloads/license-information/IEEE.html},
	shorttitle = {Dynamic {Assurance} {Cases}},
	doi = {10.1109/MC.2020.3022030},
	number = {12},
	urldate = {2025-10-21},
	journal = {Comput.},
	author = {Asaadi, Erfan and Denney, Ewen and Menzies, Jonathan and Pai, Ganesh J. and Petroff, Dimo},
	year = {2020},
	pages = {35--46},
  note = {\href{https://doi.org/10.1109/MC.2020.3022030}{doi:~10.1109/MC.2020.3022030}},
}

@inproceedings{reich_engineering_2020,
	address = {Cham},
	series = {Lecture {Notes} in {Computer} {Science}},
	title = {Engineering of {Runtime} {Safety} {Monitors} for {Cyber}-{Physical} {Systems} with {Digital} {Dependability} {Identities}},
	doi = {10.1007/978-3-030-54549-9_1},
	booktitle = {Proc. Comput. Saf., Rel., Secur.},
	publisher = {Springer International Publishing},
	author = {Reich, Jan and Schneider, Daniel and Sorokos, Ioannis and Papadopoulos, Yiannis and Kelly, Tim and Wei, Ran and Armengaud, Eric and Kaypmaz, Cem},
	editor = {Casimiro, António and Ortmeier, Frank and Bitsch, Friedemann and Ferreira, Pedro},
	year = {2020},
	pages = {3--17},
  note = {\href{https://doi.org/10.1007/978-3-030-54549-9_1}{doi:~10.1007/978-3-030-54549-9\_1}},
}

@misc{gautham_stpadriven_2022,
  title = {{{STPA-driven Multilevel Runtime Monitoring}} for {{In-time Hazard Detection}}},
  author = {Gautham, Smitha and Bakirtzis, Georgios and Will, Alexander and Jayakumar, Athira V. and Elks, Carl R.},
  year = {2022, Accessed: Nov. 01 2025},
  number = {arXiv:2204.08999},
  eprint = {2204.08999},
  primaryclass = {cs},
  publisher = {arXiv},
  doi = {10.48550/arXiv.2204.08999},
  urldate = {2025-10-01},
  archiveprefix = {arXiv},
  note = {\href{https://doi.org/10.48550/arXiv.2204.08999}{doi:~10.48550/arXiv.2204.08999}},
}

@book{leveson_engineering_2011,
  author    = {Nancy Leveson},
  title     = {{{Engineering a Safer World: Systems Thinking Applied to Safety}}},
  year = {2011},
  publisher = {MIT Press},
  doi       = {10.7551/mitpress/8179.001.0001},
  note = {\href{https://doi.org/10.7551/mitpress/8179.001.0001}{doi:~10.7551/mitpress/8179.001.0001}},
}

@article{perez_monitoring_2022,
	title = {Monitoring {ROS2}: from {Requirements} to {Autonomous} {Robots}},
	volume = {371},
	shorttitle = {Monitoring {ROS2}},
	doi = {10.4204/EPTCS.371.15},
	urldate = {2025-10-13},
	journal = {Electron. Proc. Theor. Comput. Sci.},
	author = {Perez, Ivan and Mavridou, Anastasia and Pressburger, Tom and Will, Alexander and Martin, Patrick J.},
	year = {2022},
	pages = {208--216},
	note = {
	  arXiv:2209.14030 [cs],
	  \href{https://doi.org/10.4204/EPTCS.371.15}{doi:~10.4204/EPTCS.371.15}
	},
}

@article{comella-dorda_survey_2000,
	title = {A {Survey} of {Legacy} {System} {Modernization} {Approaches}},
	url = {https://apps.dtic.mil/sti/html/tr/ADA377453/},
	language = {en},
	urldate = {2025-11-10},
	author = {Comella-dorda, Santiago and Wallnau, Kurt and Seacord, Robert C. and Robert, John},
	month = apr,
	year = {2000},
	note = {Number: CMUSEI2000TN003}
}

@inproceedings{loba_showcasing_2024,
  title = {Showcasing {{Automated Vehicle Prototypes}}: {{A Collaborative Release Process}} to {{Manage}} and {{Communicate Risk}}},
  shorttitle = {Showcasing {{Automated Vehicle Prototypes}}},
  booktitle = {Proc. 2024 IEEE 27th Int. Conf. Intell. Transp. Syst. (ITSC)},
  author = {Loba, Marvin and Graubohm, Robert and Maurer, Markus},
  year = {2024},
  pages = {3425--3432},
  doi = {10.1109/ITSC58415.2024.10919548},
  note = {\href{https://doi.org/10.1109/ITSC58415.2024.10919548}{doi:~10.1109/ITSC58415.2024.10919548}},
}

@book{wasson_system_2015,
  title = {System Engineering Analysis, Design, and Development: {{Concepts}}, Principles and Practices},
  author = {Wasson, Charles S},
  year = {2015},
  publisher = {John Wiley \& Sons}
}

@article{stolte_taxonomy_2021,
 author = {Stolte, Torben and Ackermann, Stefan and Graubohm, Robert and Jatzkowski, Inga and Klamann, Bjorn and Winner, Hermann and Maurer, Markus},
 journal = {IEEE Trans. Intell. Vehicles},
 title = {A {Taxonomy} to {Unify} {Fault} {Tolerance} {Regimes} for {Automotive} {Systems}: {Defining} {Fail}-{Operational}, {Fail}-{Degraded}, and {Fail}-{Safe}},
  year = {2021},
}

@misc{international_organization_for_standardization_iso_2018,
 author = {{International Organization for Standardization}},
 publisher = {ISO},
 title = {{ISO} 26262 {Road} {Vehicles} - {Functional} {Safety}},
  year = {2018, Accessed: Oct. 29 2025},
 urldate = {2025-08-21},
}

@inproceedings{graubohm_systematic_2017,
  title={Systematic design of automated driving functions considering functional safety aspects},
  author={Graubohm, Robert and Stolte, Torben and Bagschik, Gerrit and Reschka, Andreas and Maurer, Markus},
  booktitle = {Proc. 8. Tagung FAS},
  year = {2017},
}

@inproceedings{Maurer2002,
author = {Maurer, M. and Wörsdörfer, K.-F.},
title = {Unfallschwereminderung durch {Fahrerassistenzsysteme} mit maschineller {Wahrnehmung} -- {Potentiale} und {Risiken}},
booktitle = {Proc. Seminar Fahrerassistenzsysteme und aktive Sicherheit},
type = {Presentation},
organization = {Haus der Technik},
address = {Essen},
year = {2002},
note = {Presentation}
}

@incollection{kurtev_runtime_2017,
  title = {Runtime {{Monitoring Based}} on {{Interface Specifications}}},
  booktitle = {{{ModelEd}}, {{TestEd}}, {{TrustEd}}},
  author = {Kurtev, Ivan and Hooman, Jozef and Schuts, Mathijs},
  editor = {Katoen, Joost-Pieter and Langerak, Rom and Rensink, Arend},
  year = {2017},
  volume = {10500},
  pages = {335--356},
  publisher = {Springer International Publishing},
  address = {Cham},
  doi = {10.1007/978-3-319-68270-9_17},
  urldate = {2025-10-01},
  copyright = {http://www.springer.com/tdm},
  note = {\href{https://doi.org/10.1007/978-3-319-68270-9_17}{doi:~10.1007/978-3-319-68270-9\_17}},
}

@article{davies_interface_2020,
	title = {Interface {Management} -- the {Neglected} {Orphan} of {Systems} {Engineering}},
	volume = {30},
	doi = {10.1002/j.2334-5837.2020.00752.x},
	number = {1},
	urldate = {2025-10-07},
	journal = {INCOSE Int. Symp.},
	author = {Davies, Paul},
	year = {2020},
	pages = {747--756},
  note = {\href{https://doi.org/10.1002/j.2334-5837.2020.00752.x}{doi:~10.1002/j.2334-5837.2020.00752.x}},
}

@inproceedings{kampmann_asoa_2022,
	title = {{ASOA} -- {A} {Dynamic} {Software} {Architecture} for {Software}-defined {Vehicles}},
	copyright = {All rights reserved},
	booktitle = {Proc. 31st Aachen Colloq. Sustain. Mobility 2022},
	author = {Kampmann, Alexandru and Mokhtarian, Armin and Kowalewski, Stefan and Alrifaee, Bassam},
	year = {2022},
}

@phdthesis{klamann_ansaetze_2024,
	address = {Darmstadt},
	type = {Dissertation},
	title = {Ansätze für eine modulare {Absicherung} hochautomatisierter {Fahrzeuge}},
	urldate = {2025-11-03},
	school = {Technische Universität Darmstadt},
	author = {Klamann, Björn},
	year = {2024},
	doi = {10.26083/tuprints-00027315},
  note = {\href{https://doi.org/10.26083/tuprints-00027315}{doi:~10.26083/tuprints-00027315}},
}

@inproceedings{schubert_prototypical_2024,
  title={{A Prototypical Expert-Driven Approach Towards Capability-Based Monitoring of Automated Driving Systems}},
  author={Schubert, Richard and Kaufmann, Cedrik and Nolte, Marcus and Maurer, Markus},
  booktitle = {Proc. 2024 IEEE 27th Int. Conf. Intell. Transp. Syst.},
  pages={1051--1058},
  year={2024},
  doi = {10.1109/ITSC58415.2024.10919913},
  note = {\href{https://doi.org/10.1109/ITSC58415.2024.10919913}{doi:~10.1109/ITSC58415.2024.10919913}},
}

@inproceedings{klar_symptom_2011,
  title = {Symptom Propagation and Transformation Analysis: {{A}} Pragmatic Model for System-Level Diagnosis of Large Automation Systems},
  shorttitle = {Symptom Propagation and Transformation Analysis},
  booktitle = {Proc. ETFA2011},
  author = {Klar, Dennis and Huhn, Michaela and Gr{\"u}hser, Jochen},
  year = {2011},
  pages = {1--9},
  doi = {10.1109/ETFA.2011.6059068},
  note = {\href{https://doi.org/10.1109/ETFA.2011.6059068}{doi:~10.1109/ETFA.2011.6059068}},
}

@article{obrien_developing_1985,
	title = {Developing ‘{Expert} {Systems}’: {Contributions} from decision support systems and judgment analysis techniques},
	volume = {15},
	shorttitle = {Developing ‘{Expert} {Systems}’},
	doi = {10.1111/j.1467-9310.1985.tb00040.x},
	number = {4},
	urldate = {2025-10-21},
	journal = {R\&D Manage.},
	author = {O'Brien, W. R.},
	year = {1985},
	pages = {293--304},
	note = {
	  \_eprint: https://onlinelibrary.wiley.com/doi/pdf/10.1111/j.1467-9310.1985.tb00040.x,
	  \href{https://doi.org/10.1111/j.1467-9310.1985.tb00040.x}{doi:~10.1111/j.1467-9310.1985.tb00040.x}
	},
}

@misc{harel_methodical_2023,
	title = {Toward {Methodical} {Discovery} and {Handling} of {Hidden} {Assumptions} in {Complex} {Systems} and {Models}},
	url = {http://arxiv.org/abs/2312.16507},
	doi = {10.48550/arXiv.2312.16507},
	urldate = {2025-10-21},
	publisher = {arXiv},
	author = {Harel, David and Aßmann, Uwe and Fournier, Fabiana and Limonad, Lior and Marron, Assaf and Szekely, Smadar},
	year = {2023, Accessed: Oct. 30 2025},
	note = {
	  \href{http://arxiv.org/abs/2312.16507}{http://arxiv.org/abs/2312.16507},
	  arXiv:2312.16507 [cs],
	  \href{https://doi.org/10.48550/arXiv.2312.16507}{doi:~10.48550/arXiv.2312.16507}
	},
}

@standard{ISO26262,
  title = {{ISO} 26262 {Road} {Vehicles} -- {Functional} {Safety} -- {Part} 1: {Vocabulary}}, 
  address = {Geneva, Switzerland},
  institution = {{International} {Organization} {for} {Standardization}},
  year = {2018}
}

@techreport{waymo_road_2017,
 author = {{Waymo}},
 note = {Report},
 title = {On {The} {Road} {To} {Fully} {Self}-{Driving} - {Waymo} {Safety} {Report}},
  year = {2017},
}

@inproceedings{bagschik_systems_2018,
  title = {A {{System}}'s {{Perspective Towards}} an {{Architecture Framework}} for {{Safe Automated Vehicles}}},
  booktitle = {Proc. 21th IEEE Int. Conf. Intell. Transp. Syst.},
  author = {Bagschik, Gerrit and Nolte, Marcus and Ernst, Susanne and Maurer, Markus},
  year = {2018},
  pages = {2438--2445},
  doi = {10.1109/ITSC.2018.8569398},
  note = {\href{https://doi.org/10.1109/ITSC.2018.8569398}{doi:~10.1109/ITSC.2018.8569398}},
}

@inproceedings{lakshminarayanan_simple_2017,
	title = {Simple and {Scalable} {Predictive} {Uncertainty} {Estimation} using {Deep} {Ensembles}},
	volume = {30},
	url = {https://proceedings.neurips.cc/paper/2017/hash/9ef2ed4b7fd2c810847ffa5fa85bce38-Abstract.html},
	urldate = {2023-03-23},
	booktitle = {Proc. Advances Neural Inf. Process. Syst.},
	author = {Lakshminarayanan, Balaji and Pritzel, Alexander and Blundell, Charles},
	year = {2017},
}

@misc{labach_survey_2019,
	title = {Survey of {Dropout} {Methods} for {Deep} {Neural} {Networks}},
	url = {http://arxiv.org/abs/1904.13310},
	doi = {10.48550/arXiv.1904.13310},
	urldate = {2023-03-23},
	publisher = {arXiv},
	author = {Labach, Alex and Salehinejad, Hojjat and Valaee, Shahrokh},
	year = {2019, Accessed: Oct. 25 2025},
	note = {
	  \href{http://arxiv.org/abs/1904.13310}{http://arxiv.org/abs/1904.13310},
	  arXiv:1904.13310 [cs],
	  \href{https://doi.org/10.48550/arXiv.1904.13310}{doi:~10.48550/arXiv.1904.13310}
	},
}

@misc{kuznietsov_methodology_2024,
  title = {Methodology for a {{Statistical Analysis}} of {{Influencing Factors}} on {{3D Object Detection Performance}}},
  author = {Kuznietsov, Anton and Schweickard, Dirk and Peters, Steven},
  year = {2024, Accessed: Oct. 25 2025},
  number = {arXiv:2411.08482},
  eprint = {2411.08482},
  publisher = {arXiv},
  doi = {10.48550/arXiv.2411.08482},
  archiveprefix = {arXiv},
  note = {\href{https://doi.org/10.48550/arXiv.2411.08482}{doi:~10.48550/arXiv.2411.08482}},
  urldate = {2025-08-21},
}

@standard{iec_60812_2018,
  title        = {{ISO/IEC 60812:2018 -- Failure Modes and Effects Analysis (FMEA and FMECA)}},
  organization = {International Electrotechnical Commission (IEC)},
  institution  = {IEC},
  number       = {IEC 60812:2018},
  edition      = {3.0},
  year         = {2018},
  month        = {August},
  pages        = {165},
  language     = {English and French},
  publisher    = {VDE Verlag GmbH},
  address      = {Berlin, Offenbach}
}

@article{thomas_stpa_2015,
title = {{STPA-based Method to Identify and Control Feature Interactions in Large Complex Systems}},
journal = {Procedia Engineering},
volume = {128},
pages = {12-14},
year = {2015},
note = {Proceedings of the 3rd European STAMP Workshop 5-6 October 2015, Amsterdam},
issn = {1877-7058},
doi = {https://doi.org/10.1016/j.proeng.2015.11.499},
url = {https://www.sciencedirect.com/science/article/pii/S187770581503859X},
author = {John Thomas and Dajiang Suo}
}

@misc{abdulkhaleq_using_2017,
  title = {Using {{STPA}} in {{Compliance}} with {{ISO}} 26262 for {{Developing}} a {{Safe Architecture}} for {{Fully Automated Vehicles}}},
  author = {Abdulkhaleq, Asim and Wagner, Stefan and Lammering, Daniel and Boehmert, Hagen and Blueher, Pierre},
  year = {2017},
  number = {arXiv:1703.03657},
  eprint = {1703.03657},
  primaryclass = {cs},
  publisher = {arXiv},
  doi = {10.48550/arXiv.1703.03657},
  urldate = {2025-10-01},
  archiveprefix = {arXiv}
}

@article{graubohm_identifikation_2023,
  title = {{Identifikation ausl{\"o}sender Umst{\"a}nde von SOTIF-Gef{\"a}hrdungen durch systemtheoretische Prozessanalyse}},
  author = {Graubohm, Robert and Loba, Marvin and Nolte, Marcus and Maurer, Markus},
  year = {2023},
  journal = {- Automatisierungstechnik},
  volume = {71},
  number = {3},
  pages = {209--218},
  publisher = {De Gruyter (O)},
  doi = {10.1515/auto-2022-0164},
  urldate = {2025-10-01},
  copyright = {De Gruyter expressly reserves the right to use all content for commercial text and data mining within the meaning of Section 44b of the German Copyright Act.},
  note = {\href{https://doi.org/10.1515/auto-2022-0164}{doi:~10.1515/auto-2022-0164}},
}

@book{kletz2018hazop,
  title={{Hazop \& Hazan: Identifying and Assessing Process Industry Hazards}},
  author={Kletz, Trevor A},
  year={2018},
  publisher={CRC Press},
  doi={10.1201/9780203752227},
  note = {\href{https://doi.org/10.1201/9780203752227}{doi:~10.1201/9780203752227}},
}

@article{sun_comparison_2021,
  title = {Comparison of the {{HAZOP}}, {{FMEA}}, {{FRAM}}, and {{STPA Methods}} for the {{Hazard Analysis}} of {{Automatic Emergency Brake Systems}}},
  author = {Sun, Liangliang and Li, Yan-Fu and Zio, Enrico},
  year = {2021},
  journal = {ASCE-ASME J Risk Uncert Engrg Sys Part B Mech Engrg},
  volume = {8},
  number = {031104},
  doi = {10.1115/1.4051940},
  urldate = {2025-10-01},
  note = {\href{https://doi.org/10.1115/1.4051940}{doi:~10.1115/1.4051940}},
}

@report{vanKempen2025autotechAgil,
  author = {van Kempen, Raphael and Geller, Christian and Hülsen, Kathrin and Eckstein, Lutz and Leuffen, Marc and Lampe, Bastian and Busch, Jean-Pierre and Wirtz, Lena and Thomsen, Fabian and Winter, Mario and Feger, Ida and Kahle, Julius and Klüner, David and Hegerath, Lucas and Molz, Marius and Hartmann, Max and Steinfurth, Felix and Will, Sebastian and Braun, Niklas and Schubert, Richard and Abel, Sebastian and Bayerlein, Lorenz and Berghöfer, Moritz and Blödel, Alexander and Kuznietsov, Anton and Bahle, Jakob and Leinen, Stefan and Lauer, Martin and Le Large, Nick and Pauls, Jan-Hendrik and Poh, Willi and Rack, Nils and Steiner, Marlon and Wang, Kaiwen and Arndt, Gideon and Schulz, Benedikt and Kaljavesi, Gemb and Brecht, David and Pfab, Florian and Diermeyer, Frank and Zhou, Xingcheng and Zhang, Jiajie and Greiner, Dan and Kallfass, Ingmar and Solomakha, Oleksandr and Afanasenko, Valentyna and Hermann, Chris and Roge, Swapnil Sunil and Buchholz, Michael and Dehler, Robin and Lunic, Vladimir and Katzenbeisser, Stefan and Püllen, Dominik and Ullrich, Lars and Graichen, Knut and Jung, Lukas and Zelle, Daniel and Woopen, Timo and Böhlen, Boris and Hannig, Claudia and Lizenberg, Viktor and Mayer, Philip and Berkel, Felix and Gotzig, Heinrich and Gemlau, Kai-Björn and Hekele, Esther and Mader, Ralph and Alfranseder, Martin and Schulik, Thomas and Lilienthal, Martin},
  title = {{autotech.agil -- Architektur und Technologien zur Orchestrierung automobiltechnischer Agilität}},
  year = {2025},
  date = {2025-10-31},
  type = {Final Report},
  institution = {Technische Informationsbibliothek (TIB)},
  address = {Hannover},
  language = {German},
  pages = {97},
  doi = {10.34657/25277},
  url = {https://oa.tib.eu/renate/handle/123456789/26260}
}

\end{document}